\documentclass[a4paper,12pt]{article}
\usepackage{graphicx}
\usepackage{latexsym}
\begin{document}
%%%%%%%%%%%%%%%%%%%%%%%%%%%%%%%%%%%%%%%%%%%%%%%%%%%%%%%%%%%%
% Initialize
%%%%%%%%%%%%%%%%%%%%%%%%%%%%%%%%%%%%%%%%%%%%%%%%%%%%%%%%%%%%
%\tightenlines
\renewcommand{\baselinestretch}{1.}
\parskip 12pt
%%%%%%%%%%%%%%%%%%%%%%%%%%%%%%%%%%%%%%%%%%%%%%%%%%%%%%%%%%%%
\newcommand{\vect}[1]{\mbox{\boldmath${#1}$}}
\newcommand{\vex}{{\vect x}}
\newcommand{\vep}{{\vect p}}
\newcommand{\vev}{{\vect v}}
\newcommand{\vebt}{{\vect \beta}}

%\pagestyle{empty}
%\draft
%@twocolumnfalse\endcsname
%%%%%%%%%%%%%%%%%%%%%%%%%%%%%%%%%%%%%%%%%%%%%%%%%%%%%%%%%%%%
%\title{ Charge diffusion constant of hot and dense hadronic matter based on
%the event generator URASiMA}

\title{Charge diffusion constant in hot and dense hadronic matter \\
 - A Hadro-molecular-dynamic calculation - }

\author{N.\ Sasaki\footnote{E-MAIL:nsasaki@butsuri.sci.hiroshima-u.ac.jp}, O.\ Miyamura\footnote{E-MAIL:miyamura@sci.hiroshima-u.ac.jp},
S.\ Muroya\footnote{E-MAIL:muroya@yukawa.kyoto-u.ac.jp}$^{\ 1)} $
and C.\ Nonaka\footnote{E-MAIL:nonaka@butsuri.sci.hiroshima-u.ac.jp} \\
{\it Department of Physics, Hiroshima University, }\\
{\it Higashi-Hiroshima 739-8526, Japan }\\
$^{1)}${\it Tokuyama Women's College, Tokuyama, 745-8511, Japan  } }

\date{{\bf Version October-3, 2000 }}
\maketitle

\begin{abstract}
We evaluate charge diffusion constant of  dense
and hot hadronic matter based on the molecular dynamical
method
%with
%the relativistic nuclear collision generator URASiMA
% (Ultra-Relativistic AA collision Simulator based on Multiple
%Scattering Algorithm).
by using
a hadronic collision generator which describes nuclear collisions
at energies $10^{1 \sim 2}$ GeV/A and satisfies detailed
balance at low temperatures ($T \leq 200 $ MeV).
For the hot and dense hadronic matter of
the temperature range, $T = 80 \sim 200 $ MeV
and baryon number density, $n_{B} = 0.16$ {\rm fm}$^{-3} \sim
0.32 $ {\rm fm}$^{-3}$, charge diffusion
constant $D$  gradually increases from 0.5 fm$c$ to 2 fm$c$  with temperature
and is almost independent of baryon number density.
Based on the obtained diffusion constant we make simple discussions on the
diffusion of charge fluctuation in ultrarelativistic nuclear collisions.
\end{abstract}

%\vspace{0.5cm}

\newpage
%\baselineskip 24pt
%%%%%%%%%%%%%%%%%%%%%%%%%%%%%%%%%%%%%%%%%%%%%%%%%%%%%%%%%%%%

\section{ Introduction} \label{intr}
%%%%%%%%%%%%%%%%%%%%%%%%%%%%%%%%%%%%%%%%%%%%%%%%%%%%%%%%%%%%

Searching for the quark gluon plasma (QGP) is one of the hottest topics in
the recent high energy nuclear/particle physics \cite{QM99}.  Many kinds of
phenomena
have been proposed as candidates for the experimental signal to find the
QGP.
Charge fluctuation belongs to one of the most promising signals.  Occurrence
of
Disoriented Chiral Condensation (DCC) at QCD phase transition would
cause the large fluctuations of the ratio of the numbers of
charged pions and neutral pions \cite{DCC}.
Difference of the fluctuation intensity between QGP phase and hadronic
phase can work as a signal of the existence of QGP \cite{Jeon, Asakawa,
Fial}.
However, even if QCD phase transition takes place and the characteristic
fluctuations of QGP or QCD phase transition are produced, such a fluctuation
can be wiped out during hadronic era before the freeze-out.
Whether the fluctuation which is caused at the phase transition
or produced in the high temperature phase
can survive
in the hadronic era
or not is the diffusion problem or the transport problem in
the hot and dense hadronic matter.
Macroscopic phenomenological equations, e.g., Navier-Stokes equation and
diffusion
equation,  enable us to describe such phenomena in a simple manner.
However, those kinds of phenomenological equations contain material
constants,
so called transport coefficients, and the dynamical calculation of the
transport coefficients is a very important and difficult problem of
statistical mechanics.

Interactions between hadrons are strong interaction which are believed to be
described by QCD.  However, in the hadronic energy region,  perturbation
does
not work for QCD and no systematic way to treat is established.
In a previous paper \cite{Baryon}, we have reported a calculation of baryon
number diffusion constant based on the relativistic collision event
generator URASiMA
(Ultra-Relativistic AA collision Simulator based on Multiple scattering
Algorithm) which can reproduce hadronic spectra of nuclear collisions
in the BNL-AGS and CERN-SPS.
Usually collision event generators are so designed as to be suitable for
the description of multiple production, and the detailed balance between the
interactions is used to be paid only little attention \cite{Kumagai}.  As
we discussed in the ref.\ \cite{Baryon}, we improved the URASiMA to recover
detailed balance in the hadronic time scale and we have succeeded to
establish stationary states of interacting hadrons
with fixed temperatures and fixed baryon number densities \cite{Sasaki2}.
In this paper
 we evaluate charge diffusion constant in the hot and dense hadronic matter
and
  investigate the diffusion of the charge fluctuation in
 the  relativistic nuclear collisions.

\section{ Statistical ensembles} \label{staphys}

In order to prepare the statistical ensembles for the hot and dense hadronic
state,
we put hadrons in the box and updated with numerical code URASiMA with
periodic boundary conditions.  Recipes for the simulation
are quite similar to the ordinary molecular dynamics.
Molecular dynamics is  one of the well established numerical methods in the
statistical physics, however, it has been developed mainly for
nonrelativistic system
where particle number is conserved.  The hot and dense hadronic system in which we
have
 interest is a fully relativistic system, and the particle production and
decay  occur naturally.  Therefore,  the existence of stationary state
itself is not apparent in this system.

URASiMA is originally designed as an event generator for the relativistic
nuclear
collisions based on the hadronic multichain model (MCM).  In the high energy
collision
experiment,  multiple production takes place and the system is thought to
expand quickly.
Therefore,  usually,  production processes play essential roles in the
relativistic event   generator but the reabsorption processes 
(reversal process of
multiple production process)
do not have been thought to be important, and sometimes have been 
neglected in the simulation
code.
However detailed balance of interactions is very important for statistical
physics and
naive application of collision event generator to the molecular dynamics in
the box
leads to the one-way conversion of the energy into particle production.  As
a result,
Hagedorn  type behavior appears, i.e., strange saturation of the temperature
occurs \cite{Sasaki1,RQMD}.

In order to recover the detailed balance, we have improved URASiMA to
contain many
resonances and changed the code so as to describe some of production
processes
occur through production and decay of the resonance.  The reversal processes
of
those processes have been naturally taken into account.  As a result, after
initial thermalization time period of about 150 fm$/c$,  detailed balances
among interactions seem to be almost kept during the simulation.
It is noted that such improvement does not spoil the descriptive power
for ultrarelativistic nuclear collisions \cite{Sasaki2}.
As we have already reported in ref.\
\cite{Baryon},  slope parameter
 $T$ of the energy distribution (Table  \ref{tab:ondo}),
$$
      \frac{dN}{d^{3}\vect{p}} = \frac{dN}{4\pi Ep dE}
      = C\exp(- E/T),
$$
became almost common value for all particles (fig.\ 1) and the population
of particles became stationary (fig.\ 2).  Therefore, we looked upon the
system
as the equilibrium state with temperature  $T$ and fixed baryon number
density.
\begin{center}
===========\\
fig.\ 1 \\
===========\\
\end{center}
\begin{center}
\begin{table}[t]
\caption{
Fitted parameter $T$ of each particles at $t$ = 150 fm/{\it c}. Baryon number
density is
normal nuclear density, $n_{B}$ = 0.156 fm$^{-3}$.
 }
  \label{tab:ondo}
\begin{tabular}{c|r|r|r|r}
\hline \hline
$\varepsilon_{tot}$ [GeV/fm$^{3}$]&  $N_{938}$ [MeV] &  $\Delta_{1232}$
[MeV]
& $\pi$ [MeV] &  $\rho_{770}$ [MeV]\\               \hline
0.313 & 131 $\pm$  5 & 122 $\pm$   3 &
132$\pm$  2 & 141  $\pm$ 5  \\ \hline
0.625
&  170  $\pm$ 6
&  159  $\pm$ 5
&  163  $\pm$ 1
&  168  $\pm$ 2 \\ \hline
0.938
&  181 $\pm$  6
&  177  $\pm$ 6
&  187  $\pm$ 1
&  190 $\pm$ 2 \\ \hline

  \end{tabular}
  \end{table}
\end{center}
\begin{center}
===========\\
fig.\ 2 \\
===========\\
\end{center}

Running URASiMA many times with the same energy and the same number of
baryons,
$N_{B}$,  in the box with volume $V$,  and taking equilibrium
configurations, we
have prepared statistical ensembles of the state with temperature $T$ and
baryon number
density $N_{B}/V$.
Throughout this paper, we assumed iso-symmetry; 
the same number of protons and neutrons are put at 
initial time.

\section{  Diffusion constant in the linear response} \label{linres}

According to Kubo's Linear Response Theory, diffusion constant $D$ is
obtained current (velocity) correlation \cite{Kubo}.  Because of the
relativistic property of the hot and dense hadronic state, we should
use $\vebt = \vep /E$ instead of usual $\vev$,
\begin{equation}
D =  \frac{1}{3}\int_{0}^{\infty}<\vebt(t)\cdot \vebt(t+t')> dt' {c^2},
\label{1fdt}
\end{equation}
with $c$ being the velocity of light. In the calculation of the
charge diffusion,
average is taken over all charged particles.  When we evaluated
baryon number diffusion constant $D_{B}$ in our previous paper,   we 
took
average over only
baryons.  If correlation of velocities damps exponentially,
\begin{equation}
<\vebt(t)\cdot \vebt(t+t')> \propto \exp (- \frac{t'}{\tau} )
\label{relax}
\end{equation}
with $\tau$ being relaxation time,  we can rewrite eq.\ (\ref{1fdt})
with the simple form as,
\begin{eqnarray}
D &=&\frac{1}{3}<\vebt(t)\cdot \vebt(t)> c^2 \tau ,\label{difcon2} \\
&=&\frac{1}{3}<\left(\frac{\vep(t)}{E(t)}\right)\cdot
\left(\frac{\vep(t)}{E(t)}\right)> c^2 \tau ,
\end{eqnarray}
with use of the relaxation time.
\begin{center}
===========\\
fig.\ 3 \\
===========\\
\end{center}
Figure 3 shows the correlations of the velocity of charged particles and
exponential damping seems very good approximation.  Diffusion constant
obtained through eq.\ (\ref{difcon2}) is shown in fig. 4.
\begin{center}
===========\\
fig.\ 4 \\
===========\\
\end{center}
\noindent
Diffusion constant increases with temperature and is almost independent of
baryon number density.  This result shows clear contrast against baryon
number diffusion in our previous paper.  Baryon number diffusion constant
shows baryon number dependence and changes with temperature only weakly.
Both calculations are almost the same but main contribution
to each transport is different and we can understand fig.\ 4 as follows:
In charge diffusion, main contribution comes from charged pions
of which mass is just comparable to the temperature of the systems.
On the other hand, the mass of the baryon is much larger than the
temperature
and
baryonic system is almost nonrelativistic.  The number of the 
baryon (baryonic charge
carrier) is determined by the  baryon number of the system, and even in very
low temperature region, baryons must exist because of the fixed baryon
number.
In the lower temperature than pion mass, pion degrees of freedoms are frozen
and electromagnetic charge and baryonic charge are both carried by baryons.
At temperature about pion mass, pion degrees of
freedom start to be melted and start to contribute to the charge transfer.
In the higher temperatures, pions dominate the charge transfer.
As a result, in the temperature lower than pion mass, both diffusion
constants are almost
the same, but in the  temperature higher than  pion mass, diffusion constant
of charge current gradually increases with temperature.

\section{Charge fluctuation in ultrarelativistic nuclear collisions}

Using the above obtained values, let us sketch the diffusion of the charge
fluctuation
in the relativistic nuclear collisions.
In the simplest static picture, the solution of diffusion equation,
\begin{equation}
\frac{\partial}{\partial t}f(\vex,t)=D \nabla ^2 f(\vex,t),   \label{dfeq}
\end{equation}
with the initial distribution,
\begin{equation}
f(\vex,t_{0})=\left(\sqrt{\frac{1}{2 \pi R_{0}^{2} }} \right)^{3} {\rm
e}^{-\frac{(\vex-\vex_{0})^2}{2 R_{0}^{2}} }
\end{equation}
is given by
\begin{equation}
f(\vex, t)=\left( \sqrt{\frac{1}{\pi (2R_{0}^{2} +4 D (t-t_{0}) )} }
\right)^{3}
{\rm e}^{-\frac{(\vex-\vex_{0} )^2}{2 R_{0}^{2} +4 D (t-t_{0}) } } .
\end{equation}
Suppose that charge fluctuation with size $R_{0}$ produced at $t=t_{0}$,
subjecting to
eq.\ (\ref{dfeq}), the fluctuation diffuses and expands to
$R(t)=\sqrt{R_{0} ^{2} + 2D(t-t_{0})  } $.  Therefore, we may regard that at
 such
time $t$ that $ 2D(t-t_{0}) = R_{0} ^{2}$,
charge fluctuation almost disappears.  If the charge fluctuation about $R_{0}=3$ fm
at
$T=T_{C}\cong $ 160 MeV is produced,  it survives only 2 fm/{\it c}
during 
hadronic
matter era.
But if charge fluctuation is produced at lower temperature about 120
MeV (Super
Cooled DCC), diffusion
constant is about 0.7 fm$c$ and the 
fluctuation with initial size of about 3 fm
will survive for 6 fm/{\it c}.  The
length of hadronic era and the chronological change of the temperature
depend on
the solution of the hydrodynamical model, but the latter case is promising
to be observed experimentally.

In order to take account of expansion of the hadronic fluid, we must use
relativistic
hydrodynamical model.  In the relativistic hydrodynamical model, charged
current
$J^{\mu}$ is given as,
\begin{equation}
J^{\mu}=f(\vex, t) U^{\mu} +J_{d}^{\mu},
\end{equation}
with $U^{\mu}$ being local four velocity \cite{Namiki,Landau}.
 $f(\vex, t)$ is charge density on the local rest frame,
$f(\vex, t) = U^{\mu} J_{\mu}$.
The diffusion current, $J_{d}^{\mu}$ is given by relativistic extension of
the
Fick's law\footnote{In the relativistic notation, $x^0 = ct$, 
and $\frac{\partial}{\partial x^{0}} = \frac{\partial}{c \partial t}$.}
\begin{equation}
J_{d}^{\mu} = \frac{D}{c}{\Delta^{\mu}}_{\nu}\partial^{\nu}f(\vex, t),
\end{equation}
where ${\Delta^{\mu}}^{\nu}=g^{\mu \nu}-U^{\mu}U^{\nu}$ is space-like
projection operator
orthogonal to $U^{\mu}$.
Putting above $J^{\mu}$ into the continuity equation, $\partial
_{\mu}J^{\mu} =0 $, we
can obtain the relativistic diffusion equation,
\begin{equation}
U^{\mu}\partial_{\mu}f(\vex, t) +f(\vex, t)\partial_{\mu}U^{\mu}
+\frac{D}{c} \partial_{\mu}{\Delta^{\mu}}_{\nu} \partial^{\nu} f(\vex, t)  = 0.
\label{rdfeq}
\end{equation}

In the case of 1+1 dimensional scaling expansion, particular solution of the
local four velocity is given as
$U^{\mu}=\frac{x^{\mu} }{c \tau}$ with  light-like coordinate, $c t=c \tau
\cosh \eta$,
 $z =c \tau \sinh \eta$ \cite{Akase}, and diffusion equation (\ref{rdfeq}) is
 rewritten as,
\begin{equation}
{\frac{\partial}{\partial \tau}} f(\eta, \tau) +\frac{1}{\tau} f(\eta,
\tau)=
{\frac{D}{c^{2} \tau ^2}}{\frac{\partial ^2}{\partial \eta ^2}}f(\eta,\tau).
\label{scal}
\end{equation}
The solution of eq.\ (\ref{scal}) with initial condition,
$ f(\eta, \tau_{0})= \sqrt{\frac{1}{2 \pi R_{0}^{2} }}
 \exp ({-\frac{ \eta ^2}{2 R_{0}^{2}} })f_0, 
$ is given by 
\begin{equation}
f(\eta, \tau) = \frac{\tau_0}{\tau}\sqrt{\frac{1}{2 \pi R(\tau)^{2} }}
 {\rm e}^{-\frac{ \eta ^2}{2 R(\tau)^{2}} }f_0  \label{sol2}
\end{equation}
with $ {R(\tau)}^{2} $ being
\footnote{Note that this is the difussion in the $\eta$ direction 
and $R$ is the size in the $\eta$ coordinate.
},
\begin{equation}
R(\tau)^{2} = R_{0}^{2} + 2 \frac{ D}{{c^2} {\tau \tau_0}} (\tau-\tau_{0}).
\label{hankei}
\end{equation}
The first factor in eq.\ (\ref{sol2}), $\frac{1}{\tau}$, is the result of
systematic
expansion of the scaling solution;  the region with fixed
$\Delta \eta$ expands as $\tau \Delta \eta$.
This rapid expansion also dominates the evolution of the fluctuation size,
$R(\tau)^{2}$; the region with fixed $\Delta \eta$ expands with $\tau$
but the diffusion effect
subject to eqs.\ (\ref{dfeq}) or ( \ref{rdfeq}) is the expansion with 
only $\sqrt{2D \tau}$.
Therefore, in $\eta$ coordinate, the effect of the diffusion becomes  
smaller and smaller with time, $\frac{ D}{{c^2} {\tau \tau_0}}$.
According to our result (fig.\ 4), diffusion constant in the hadronic matter
is at
most 2 fm$c$ and  the diffusion term in eq.\ (\ref{hankei}) becomes small
enough  already at $\tau_{0}$= several fm.  As a result, the fluctuation in
$\eta$ will
be deformed only little by the diffusion effect in the hadronic era.

\section{ Concluding remarks } \label{conrem}

We evaluated charge diffusion constant of hot and dense hadronic matter
based on the
relativistic collision event generator URASiMA.   Obtained charge diffusion
constants
increase from 0.5 fm$c$  to 2 fm$c$ in the temperature range between
80 MeV and 
200 MeV.  Based on the obtained diffusion constant, we made  rough sketches
of the
diffusion of charge fluctuation in the ultrarelativistic nuclear collisions.
For more improved discussion, we need to solve the hydrodynamical equation
coupled with the charge current conservation equation \cite{Ishii}.

\noindent
{\bf Acknowledgment}

The authors would like to thank prof.\ M.\ Namiki for his fruitful comments.
They also would like to thank Dr.\ K.\ Homma for drawing our attention
to this topic.
Discussions with Prof.\ A.\ Nakamura and Prof.\ S.\ Dat{\'e} were quite
valuable.
This work is supported by Grant-in-Aid for scientific research number
11440080 by Ministry of Education, Science, Sports and Culture,
Government of Japan (Monbusho).  Calculation has been done at Institute for
Nonlinear
Sciences and Applied Mathematics, Hiroshima University.

\newpage
\begin{figure}[tbh]
\begin{center}
\includegraphics[scale=0.73]{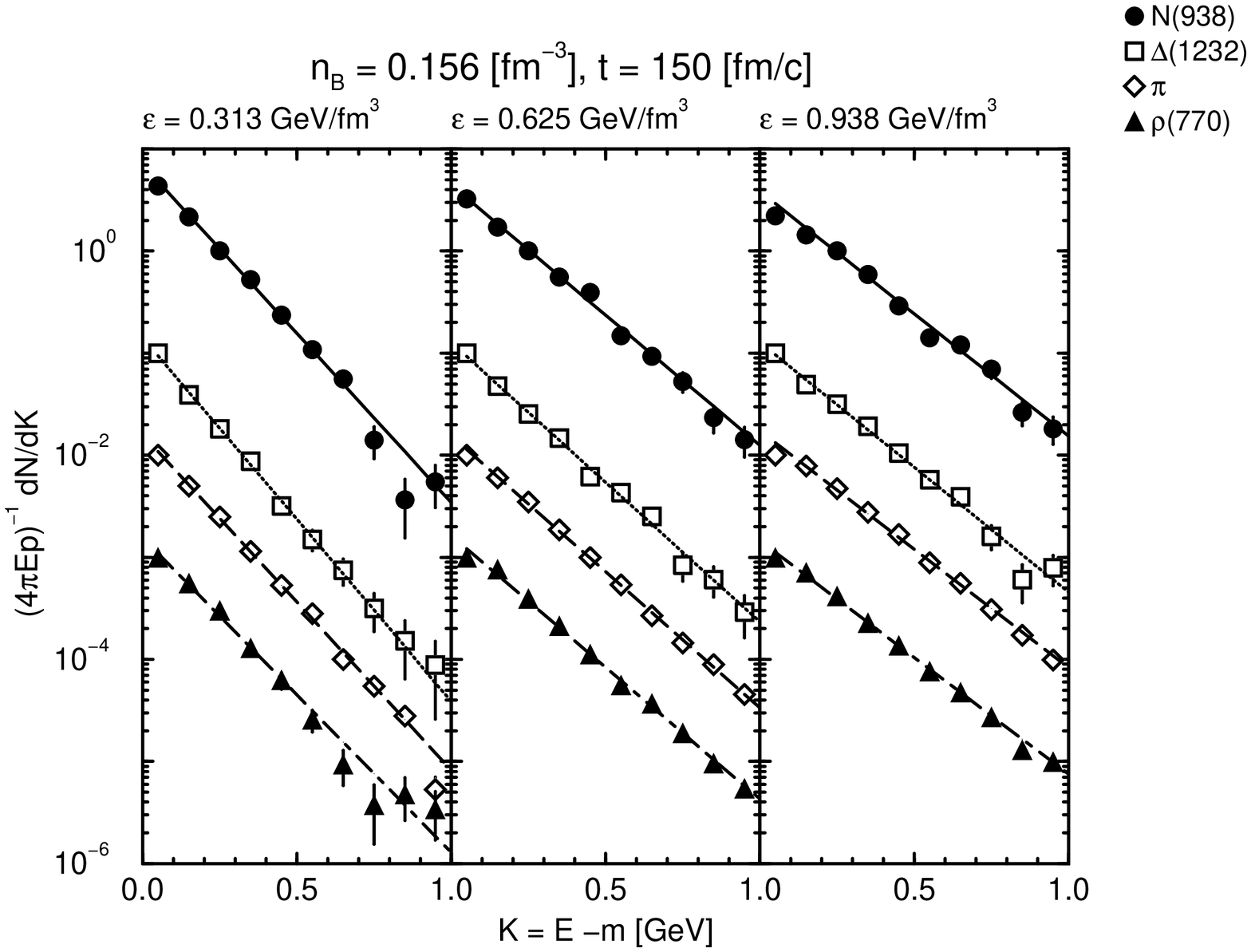} \\
\caption{Energy distributions of particles at $t=150$ fm/{\it c}. 
Lines stand for
the results of fitting with thermal distribution.
Normalizations of the data are arbitrary.}
\end{center}
\end{figure}
\newpage
\begin{figure}[tbh]
\begin{center}
\includegraphics[scale=0.73]{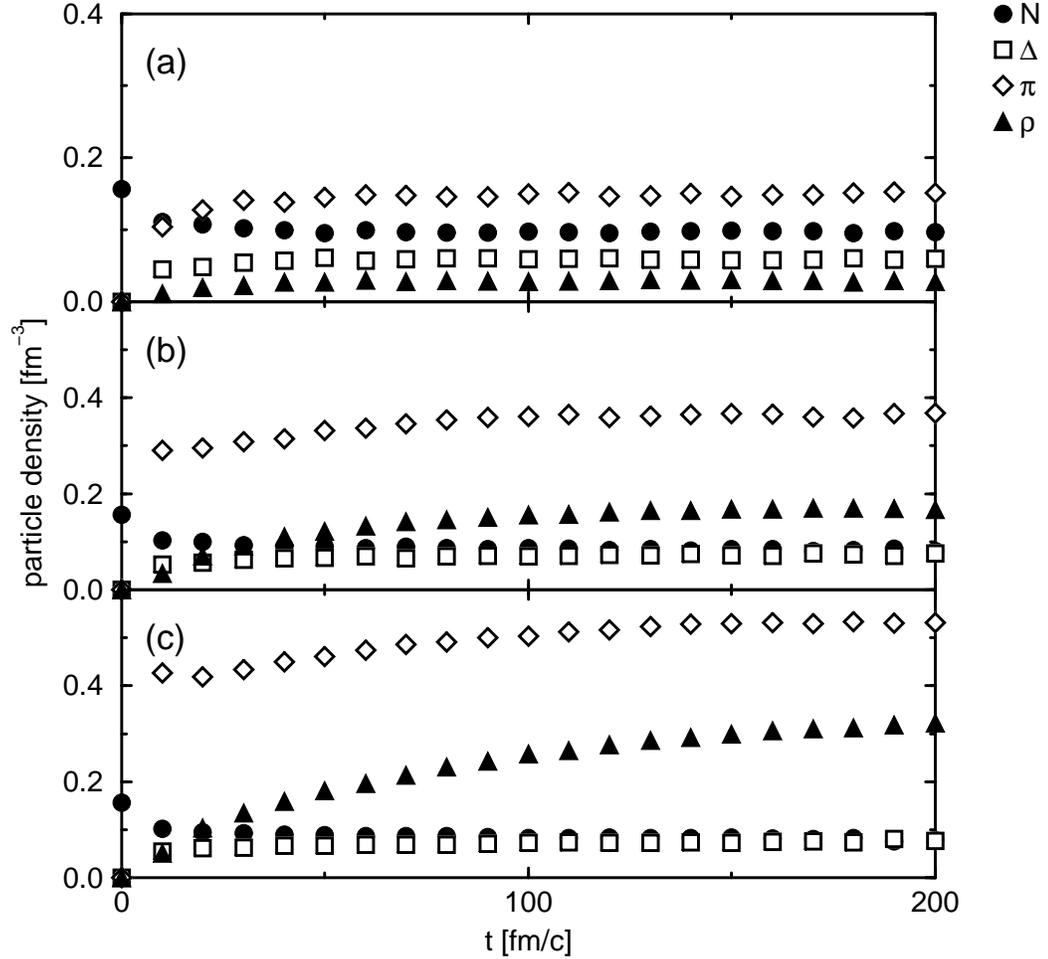} \\
\caption{Particle density of time. The population of the particles in later
than $t=150$
 fm/c seem to be stationary. In each case, baryon number density is
$n_{B}=0.156$ 
 fm$^{-3}$, and energy
density for (a) is 0.313 GeV/fm$^{3}$, (b) is 0.625 GeV/fm$^{3}$ and (c)
is 0.938  GeV/fm$^{3}$, respectively.
}
\end{center}
\end{figure}
  \newpage
\begin{figure}[tbh]
\begin{center}
\includegraphics[scale=0.75]{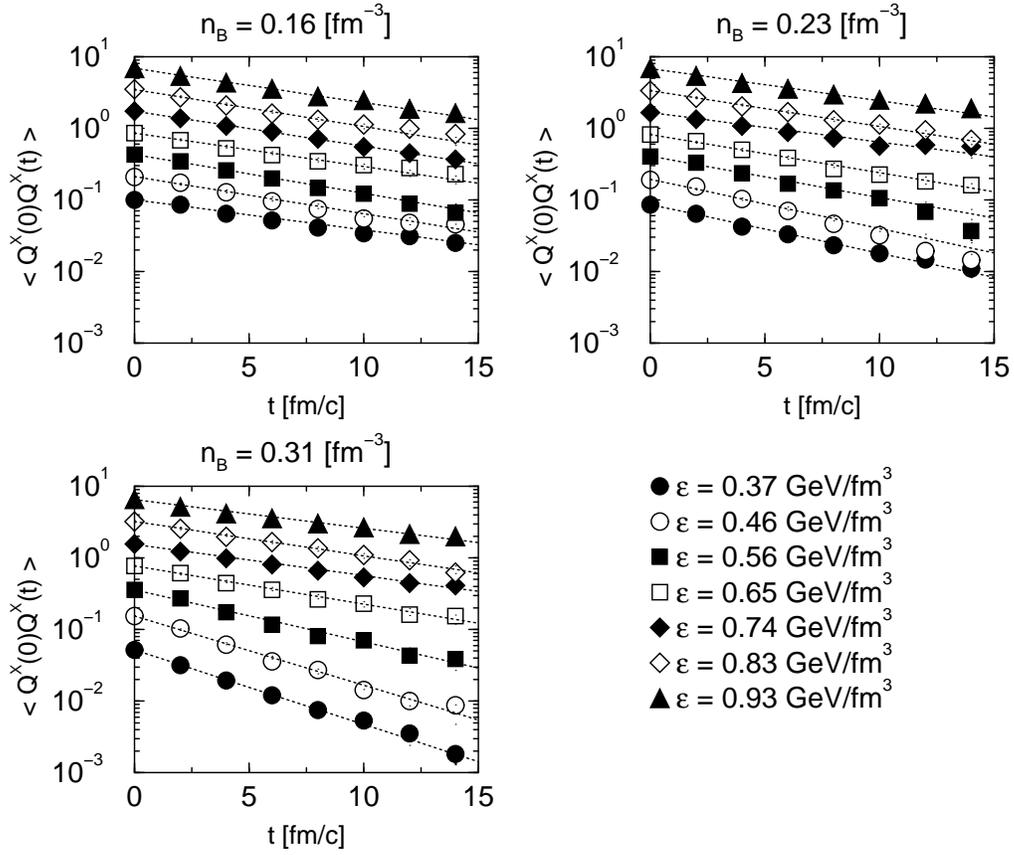} \\
\caption{Velocity correlation of the charged particles as a function of
time.
    Lines correspond to the fitted results by exponential function eq.\ (2).
    Normalizations of the data are arbitrary.}
\end{center}
\end{figure}
\newpage
\begin{figure}[tbh]
\begin{center}
\includegraphics[scale=0.75]{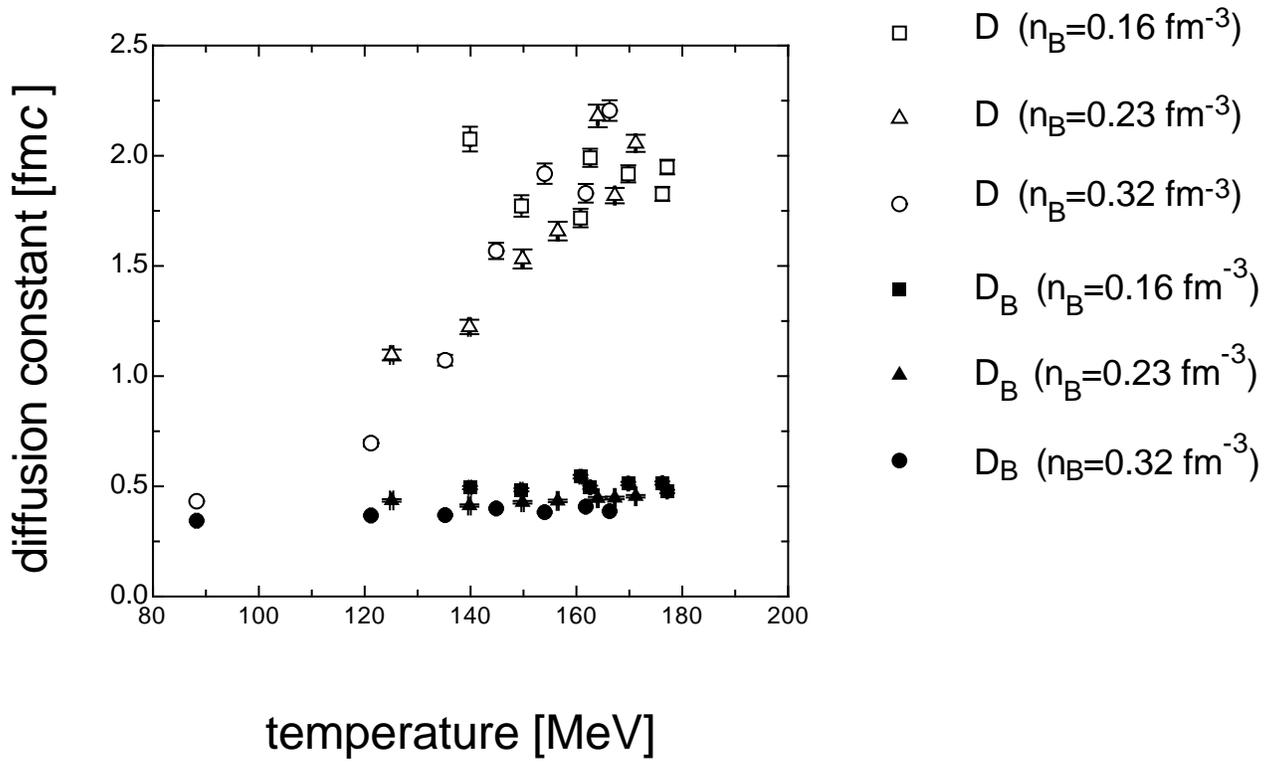} \\
\caption{Diffusion constant $D$ and baryon diffusion constant $D_{B}$. }
\end{center}
\end{figure}

\end{document}